
\input harvmac.tex
\noblackbox

\def\csc{closed string channel}
\def\osc{open string channel}

\def\ket#1{|#1\rangle}
\def\bra#1{\langle#1|}

\lref\smat{J. Polchinski, Nucl. Phys. {\bf B362} (1991) 125;
G. Moore, Nucl. Phys. {\bf B368} (1992) 557;
K. Demeterfi, A. Jevicki and J. Rodrigues, Nucl. Phys. {\bf B362} (1991)
173; G. Mandal, A. Sengupta and S. Wadia,
Mod. Phys. Lett. {\bf A6} (1991) 1465.}
\lref\discstat{V. G. Kac, in {\sl Group Theoretical
Methods in Physics}, Lecture Notes in Physics, vol. 94
(Springer-Verlag, 1979).}
\lref\KP{I. R. Klebanov and A. M. Polyakov,
Mod. Phys. Lett. {\bf A6} (1991) 3273.}
\lref\BPZ{A. Belavin, A. Polyakov and A. Zamolodchikov,
Nucl. Phys. {\bf B241} (1984) 333.  }
\lref\hof {D. R. Hofstadter, Phys. Rev. {\bf B14} (1976) 2239.}
\lref\wannier{G. H. Wannier, Phys. Status Solidi  {\bf B88} (1978) 757.}
\lref\cgcdef{C. G. Callan and D. Freed,``Phase Diagram of the Dissipative
Hofstadter Model'', Princeton Preprint PUPT-1291, June 1991.}
\lref\dissqm{A. O. Caldeira and A. J. Leggett, Physica  {\bf 121A}(1983) 587;
Phys. Rev. Lett. {\bf 46} (1981) 211; Ann. of Phys. {\bf 149} (1983) 374.}
\lref\osdqm{C. G. Callan, L. Thorlacius, Nucl. Phys. {\bf B329} (1990) 117.}
\lref\cardyone{J. L. Cardy, J. Phys. {\bf A14} (1981) 1407.}
\lref\clny{C. G. Callan, C. Lovelace, C. R. Nappi, and S. A. Yost,
Nucl. Phys. {\bf B293} (1987) 83; Nucl. Phys. {\bf B308} (1988) 221.}
\lref\CTrivi{C. G. Callan, L. Thorlacius, Nucl.  Phys. {\bf B319} (1989) 133.}
\lref\fisher{M. P. A. Fisher and W. Zwerger, Phys. Rev. {\bf B32} (1985) 6190.}
\lref\klebanov{I. Klebanov and L. Susskind, Phys. Lett. {\bf B200} (1988) 446.}
\lref\ghm{F. Guinea, V. Hakim and A. Muramatsu,Phys. Rev. Lett.  {\bf 54}
(1985) 263.}
\lref\afflud{I. Affleck and A. Ludwig, Phys. Rev. Lett {\bf 67} (1991) 161.}
\lref\tseyt{E. Fradkin and A. Tseytlin, Phys. Lett. {\bf 163B} (1985) 123.}
\lref\abou{A. Abouelsaood, C. G. Callan, C. R. Nappi and S. A. Yost,
Nucl. Phys. {\bf B280} (1987) 599.}
\lref\kondo{I. Affleck and A. Ludwig, Nucl.Phys. i{\bf B352} (1991) 849.}
\lref\caldas{C. G. Callan and S. Das,  Phys. Rev. Lett {\bf 51} (1983) 1155.}
\lref\cardytwo{J. L. Cardy, Nucl. Phys. {\bf B324} (1989) 581. }
\lref\isqm{C. G. Callan, ``Dissipative Quantum Mechanics in Particle Physics",
Princeton preprint PUPT-1350, in ``Proceedings of the Fourth International
Conference on Quantum Mechanics in the Light of New Technology",
S. Kurihara ed., Japan Physical Society (1992). }
\lref\gsw{ M. B. Green, J. H. Schwarz, and E. Witten, ``Superstring Theory'',
Cambridge University Press (1987).}
\lref\freed{D. Freed, ``Contact Terms and Duality Symmetry in the Critical
Dissipative Hofstadter Model'', MIT preprint MIT-CTP-2170, Mar 1993,
hep-th 9304006.}
\lref\ishi{N. Ishibashi and T. Onogi , Mod.Phys.Lett. {\bf A41} (1989) 161.  }


\Title{\vbox{\baselineskip12pt
\hbox{PUPT-1432}\hbox{IASSNS-HEP-93/78}\hbox{hepth@xxx/9311092}}}
{Exact C=1 Boundary Conformal Field Theories}
\centerline{Curtis G.
Callan\footnote{$^\diamondsuit$}{callan@puhep1.princeton.edu}
\footnote{$^\clubsuit$}{On leave from Princeton University.}}
\centerline{\it School of Natural Sciences, Institute for Advanced Study}
\centerline{\it Princeton, NJ 08544}
\centerline{and}
\vskip .2in
\centerline{Igor R. Klebanov\footnote{$^{\spadesuit}$}
{klebanov@puhep1.princeton.edu}}
\centerline{\it Department of Physics, Princeton University}
\centerline{\it Princeton, NJ 08544}
\vskip .3in
\centerline{\bf Abstract}
We present a solution of the problem of a free massless scalar field
on the half line interacting through a periodic potential on the boundary.
For a critical value of the period, this system is a conformal
field theory with a non-trivial and explicitly calculable S-matrix for
scattering
from the boundary. Unlike all other exactly solvable conformal field theories,
it is non-rational ({\it i.e.} has infinitely many primary fields). It
describes
the critical behavior of a number of condensed matter systems, including
dissipative quantum mechanics and of barriers in ``quantum wires''.
\smallskip

\Date{11/93}

\newsec{Boundary Conformal Field Theory}

Conformal field theory is usually defined on a two-dimensional manifold
without boundaries (the simplest case being the plane). It can also be
defined on manifolds with boundaries (like the disk or strip),
provided that appropriate boundary conditions are imposed \cardyone.
The Dirichlet and Neumann boundary conditions on scalar worldsheet fields
are familiar, if trivial, examples. Non-trivial conformal boundary
conditions arise from the interaction of boundary degrees of freedom with
worldsheet fields. A wide range of systems, including open string theory
\refs{\clny,\abou,\tseyt}, monopole catalysis \caldas, the Kondo problem
\kondo, dissipative quantum mechanics \refs{\dissqm,\osdqm} and junctions
in quantum wires \isqm\ can be described this way.

The technology for dealing with boundary conformal field theory is
easily stated \cardytwo:
Consider a bulk conformal field theory {\cal C} confined
to a strip of width $L$ with boundary conditions A and B on the two ends.
This theory has a partition function
$Z_{open}^{AB}=tr(e^{-TL_0^{AB}})$ where $T$ is the time interval and
$L_0^{AB}$ is the open string Hamiltonian. If the boundary conditions
are conformal, the partition function will be a sum
$Z_{open}^{AB}=\sum n_h\chi_h(e^{-2\pi T/l})$ over Virasoro characters of the
open string primary fields (the $h$ are the highest weights and the
$n_h$ are the integer multiplicities of the characters). The partition function
can
also be computed as the amplitude for {\it closed} string propagation
between states $\ket{A}$ and $\ket{B}$ of the bulk {\it closed} string
created by the boundary conditions $A$ and $B$: $Z_{closed}^{AB}=
\bra{A}e^{-l(L_0+\widetilde L_0)}\ket{B}$, where $L_0$ and $\widetilde L_0$ are
the left- and right-moving closed string Hamiltonians.
For the theory as a whole to be conformal, the boundary states must
satisfy a reparametrization invariance condition
$(L_n-\widetilde L_{-n})\ket{A}=0$ \clny\ which implies that each primary field
contributes to $\ket{A}$ a piece $C^A_h\sum_n\ket{n}\ket{\widetilde n}$,
where the sum is over all the states of the Virasoro module and $C^A_h$ is a
coefficient to be determined \ishi. This gives a different expansion
of the partition function in terms of Virasoro characters:
$Z_{closed}^{AB}=\sum_h C^A_hC^B_h\chi_h(e^{-2\pi l/T})$.
The dynamical problem
is to find the specific primary fields $\phi_h$ appearing in both the open and
closed string expansions along with their multiplicities and weights.

The two expansions must of course be identical
under the ``modular transformation'' $e^{-2\pi T/l}\to e^{-2\pi l/T}$
between open and closed string variables. This consistency condition
is often enough to explicitly determine the (finitely many) boundary states
of a rational conformal field theory such as the WZW theory (which underlies
the Kondo model \afflud). Non-rational conformal field theories are much harder
to
deal with since they have an infinite number of primary fields and, presumably,
boundary states. In this Letter we will present an approach to
the solution of what must be
the simplest such theory: a single free scalar field interacting only via a
boundary
periodic potential. Our solution is part conjecture, but its
combination of richness
and simplicity convinces us that it is exact. As mentioned above, there are
several interesting condensed matter systems to which our results apply.

\newsec{The Free Scalar With Periodic Boundary Potential}

We will study free massless
scalar field theory on the interval $0<\sigma<l$.
A dynamical boundary condition at $\sigma=0$ is
imposed by including a potential
term in the otherwise free Lagrangian:
\eqn\lag{L =
{1\over 8\pi}\int_0^l d\sigma (\partial_\mu X)^2-
{1\over 2\epsilon} (g e^{iX(0)/\sqrt 2}+
{\bar g} e^{-iX(0)/\sqrt 2})}
where $\epsilon$ is the short-distance cutoff and $g$ is a complex renormalized
potential strength. To control infrared problems we impose a Dirchlet boundary
condition, $X(l)$=0, at $\sigma=l$,
but we eventually want to focus on the physics at the $\sigma=0$ boundary.
The potential induces a perturbation away from the (manifestly conformal)
free scalar field subject to the Neumann (Dirichlet) boundary
condition on the left (right) end of the interval. The specific potential of
Eqn.~\lag\ was chosen because it has
boundary scaling dimension one and induces a
marginal perturbation away from the conformal fixed point \fisher.
We will show that it is
in fact {\it exactly} marginal and induces a conformal boundary condition for
{\it all} values of $g$.

For the subsequent analysis it will be helpful to recall some facts about the
primary
fields of conventional $c=1$ CFT: Modulo some subtleties which do not affect
our
application, there is a continuum of holomorphic primary fields
$e^{ikX(z)}$ of weights $h=k^2/2$ (and corresponding antiholomorphic fields).
The associated Virasoro characters are $\chi_k(q)=q^{k^2/2}/f(q)$, where
$f(q)=\prod_{n=1}^\infty (1-q^n)$. For special values of the ``momentum''
$k$, some descendant states have vanishing norm and new primaries,
the famous discrete states appear \discstat.
They are organized in $SU(2)$ multiplets
of spin $J=0, {1\over 2}, 1,\ldots $. The $(J, m)$ primary, $\psi_{(J, m)}$,
has weight $h=J^2$ and a Virasoro character which turns out to be
${\chi}_{(J,m)}(q)=(q^{J^2}-q^{(J+1)^2})/f(q)$.
There is an explicit representation for $\psi_{(J,m)}$ \KP,
\eqn\ds{
\psi_{(J, m)}(0)\sim \left [\oint {dz\over 2\pi i} e^{-i\sqrt 2 X (z)}
\right ]^{J-m} e^{iJ\sqrt 2 X (0)},}
which shows that it is a polynomial in $\partial X$, $\partial^2 X$,
etc. times a zero mode piece $e^{im\sqrt 2 X}$. The lowest-lying fields,
$\psi_{1/2,\pm 1/2}\sim e^{\pm i X/\sqrt 2}$, are precisely the terms
appearing in the boundary potential in \lag. Since products of
$\psi_{(J, m)}$ fuse to other $\psi_{(J, m)}$ by an $SU(2)$
fusion algebra, the operators appearing in a perturbation expansion in the
boundary potential should be spanned only by the discrete states.
This strongly suggests that the exact boundary states are sums
over the Virasoro modules of the discrete state primary fields $\psi_{(J, m)}$.

Let us first check that the conjecture is true when the potential vanishes
(free field with one Neumann and one Dirichlet boundary condition). The
partition function is easily found to be
$$ Z_0= w^{1/48}\prod_{n=1}^\infty {1\over 1-w^{n-{1\over 2}}}=
{w^{-1/24}\over f(w)} \sum_{j=0}^\infty w^{(j+1/2)^2/4}\ ,
$$
where $w=e^{-2\pi T/l}$. This shows that the energy levels of this open string
organize themselves into a set of Virasoro
modules with highest weights $h_j=\pi (j+{1\over 2})^2/(2l)$. Using standard
technology, we can re-express $Z_0$ in terms of the closed string variable
$q^2=e^{-2\pi l/T}=e^{-2\pi\tau}$, with the result
\eqn\partz{ Z_0= {(q^2)^{-1/24}\over f(q^2)}\sum_{n=-\infty}^\infty (-1)^n q^{2
n^2}
={(q^2)^{-1/24}\over f(q^2)}\theta_4(0|2 i\tau)\ ,}
where the theta functions are defined using the conventions of \gsw.
Because of the discrete state subtlety, it is  not so obvious how to read off
the Virasoro modules which propagate in the \csc. However,
a little experimentation shows that the contributions
of the discrete modules to the boundary states are
\eqn\bound{\eqalign{&\ket{B_N}=\sum_{J=0}^\infty  (-1)^J |J, 0\gg \ ,\cr
&\ket{B_D}=\sum_{J=0}^\infty\sum_{m=-J}^J |J, m\gg \ ,}
}
where $|J, m\gg $ is the module associated with $\psi_{(J, m)}$
($\ket{B_D}$ also receives contributions from the continuum states $e^{ikX}$,
although they have no effect on our considerations).
Eqs. \bound\ reproduce \partz\ and all partition functions arising from
other combinations of Neumann and Dirichlet boundary conditions. To see this
one just expands the basic boundary state formula
$Z_0=\bra{B_N}e^{-l(L_0+\widetilde L_0)}\ket{B_D}$ with the
help of the discrete state character formula
$$\ll J, m|e^{-l (L_0+\widetilde L_0)}|J', m^\prime\gg =
	\delta_{J J'}\delta_{m m^\prime}
{(q^2)^{-1/24}\over f(q^2)} \left [ q^{2J^2}- q^{2(J+1)^2}\right ] .
$$

\newsec{Some Perturbative Results}

We now want to turn on the potential and
expand the partition functions in powers
of $g$ and $\bar g$. Since only one boundary is dynamical, the only new
element we need is the massless scalar propagator between two
points on the same boundary of a cylinder of length $\tau=l/T$ and
circumference $1$. This standard object is expressed in terms of theta
functions as
\eqn\prop{\vev{X(t_1) X(t_2)}_{\sigma=0}
=-2 \log {\theta_1^2 \bigl (t_1-t_2 | 2i\tau\bigr)
\over \theta_4^2 \bigl (t_1-t_2 | 2i\tau\bigr) }}
Expanding the partition function to first order gives
\eqn\zone{\eqalign{
Z_1 = & -Z_0 {(g+\bar g)\over 2\epsilon}\int_0^1
	\langle e^{iX(t)/\sqrt{2}}\rangle \cr
	=& -Z_0~{(g+\bar g)\over 2\epsilon}e^{-\vev{X(0) X(\epsilon)}/4}=
	-Z_0~{(g+\bar g)\over 2}~
{\theta^\prime_1(0|2i\tau)\over\theta_4(0|2i\tau)}\ .}}
Note that we have eliminated the divergence of this amplitude
by regulating the integrand (by point-splitting of the coincident-point
propagators) and then by renormalizing (by taking the potential strength
proportional to $1/\epsilon$).
The net first-order result, expressed in \csc\ variables, is
\eqn\zoneexp{ Z_1=\pi (g+\bar g) {(q^2)^{-1/24}\over f(q^2)}
\sum_{n=0}^\infty (-1)^{n+1} (2n+1) q^{(2n+1)^2/2}.}
The powers of $q^2$ which appear in the sum correspond to the weights of the
discrete states $\psi_{(J,\pm1/2)}$ for
all possible half-integer $J$ and we can
find a corrected $\ket{B_N}$, containing such states, which reproduces
\zoneexp.
The dual transformation to the \osc\ gives
$$ Z_1 =-{\pi T\over 4l}(g+\bar g)
{w^{-1/24}\over f(w)} \sum_{j=0}^\infty (-1)^j (2j+1) w^{(j+1/2)^2/4}\ ,
$$
This can be interpreted as a shift of the highest weights of
the Virasoro modules appearing in $Z_0$, with the
$j$-th module being shifted by $(-1)^j (2j+1)\pi (g+\bar g)/(4l)$. The main
point here is that the perturbation causes all the energy levels of any given
Virasoro module to have a common energy shift, which they must if
the perturbation in \lag\ is truly conformal.

\newsec{Conjectured Exact Solution}

Now consider the expansion of $Z(g, \bar g)$ to higher orders. A new
power law divergence, contributing to a shift of the open string vacuum energy
and arising from the collision of an $e^{iX/\sqrt 2}$ with an $e^{-iX/\sqrt 2}$
insertion first appears in second order. It turns out to be possible to
subtract
the divergence in this, and all higher, orders by a simple principal value
prescription. The second-order terms ({\it i.e.} the $g^2$, $\bar g^2$ and
$g\bar g$ terms) organize nicely into
$$Z_2=\pi^2 (g+\bar g)^2 {(q^2)^{-1/24}\over f(q^2)}
\sum_{n=1}^\infty (-1)^{n+1} q^{2 n^2} n^2 .
$$
To this order,
the partition function $Z_0+Z_1+Z_2$, when reexpressed in \osc\ variables,
can once again be interpreted as a sum over open string Virasoro modules with
shifted highest weights.
This is a new piece of evidence that the theory specified by \lag\ is exactly
conformal.

We have carried out the expansion of $Z$ to fourth order and continue to find
results consistent with exact conformal invariance. We have even found a
general expression for the partition functions which we believe to summarize
the behavior of the theory to all orders: Everything we know is consistent
with the net effect of the boundary interaction being a shift of the
highest weights of the open string Virasoro modules by a universal, coupling
constant dependent, shift function. We claim that the exact weight of the
$j$-th open string module has the form
$$h_j ={\pi \left(j+{1\over 2}+(-1)^j {\Delta(g,\bar g)/ \pi}\right)^2
\over 2l}$$
so that the \osc\ partition function has the form
\eqn\partopen{
Z(g, \bar g)=
{w^{-1/24}\over f(w)} \sum_{j=0}^\infty w^{[j+1/2+
(-1)^j \Delta(g,\bar g)/\pi]^2/4}.}
Calculations out to fourth order are all consitent with \partopen\ with
\eqn\phase{ \Delta(g,\bar g)={\pi\over 2}(g+\bar g)+
{\pi^3\over 48}(g^3+\bar g^3-3 g^2\bar g-3 g\bar g^2)+\ldots~. }

When \partopen\ is transformed to \csc\ variables, we obtain
\eqn\partclosed{Z(g,\bar g) =
{(q^2)^{-1/24}\over f(q^2)}
\left (1+2\sum_{n=1}^\infty q^{n^2/2} \cos \left [{n\pi\over 2}+n\Delta
(g,\bar g)\right ]\right)~.}
The remarkable thing about this expression is that, for any value of $\Delta$,
it involves {\it only} the weights of the discrete states. Actually, some
further conditions have to be met in order for it to be possible to construct
\partclosed\ by sandwiching the closed string propagator between discrete
state boundary states. It is easy to see that on expanding \partclosed\ in
powers of $g$ and $\bar g$, the term of order $g^k \bar g^l$ must come from
$(J,m)$ modules with $m={k-l\over 2}$. Since such modules have
$J\ge |{k-l\over 2}|$, and since the discrete state weights are $J^2$,
the $q^2$ expansion of the $g^k \bar g^l$ term must begin at $q^{(k-l)^2/2}$.
If these conditions are met one can find an explicit
expansion of the dynamical boundary state of the form
\eqn\dynbs{
\ket{B~(g,\bar g)}=\sum_{J=0}^\infty \sum_{m=-J}^{J}
C_{Jm}(g,\bar g) |J, m\gg }
where the $C_{Jm}$ are expressed in terms of operations carried out on
$\Delta$. Low order perturbation theory calculations imply that
theory,
$$ C_{Jm}=(-1)^J \delta_{m, 0}+{\pi\over 2}
(-1)^{(2J+1)/2} (2J+1)
\left (g\delta_{m, 1/2}+\bar g\delta_{m, -1/2}\right)+\ldots~.
$$
(Some notational license has been taken in that the first(second) term is
present only for integer (half-integer) $J$.)

Although we don't have complete knowledge
of $\Delta$, some interesting things can be said about it. The consistency
conditions fix all $g^k$ or $\bar g^k$ terms in
the expansion of $\Delta(g, \bar g)$ in terms of the leading $O(g)$ term.
This turns out to imply that $\Delta(g, 0)=\arcsin(\pi g/2)$.
Other information on $\Delta(g,\bar g)$ comes from
considering the strong potential limit, {\it e.g.} $g=\bar g \to \infty$.
It is physically clear that, in this limit the boundary state should reduce
to a sum over the Dirichlet boundary states localized at the minima of
$\cos (X/\sqrt 2)$.
Similarly, for $g\to -\infty$ the end of the string is localized
at the maxima of $\cos (X/\sqrt 2)$. Indeed, the correct partition
function in these limits results from Eqn.~\partopen\ if
$\Delta(g\to \infty)=-\Delta(g\to -\infty)=\pi/2$.

\newsec{The Exact $S$-Matrix}

The universal function $\Delta(g,\bar g)$ should implicitly contain
all the physical information about the theory. A particularly
interesting set of questions arises in calculation of the reflection
$S$-matrix from the dynamical boundary at $\sigma=0$, which is
well-defined for a semi-infinite string. The $S$-matrix is determined
by correlation functions of operators $\partial X$ and $\bar \partial X$
on a Euclidean half-plane with the interaction of Eq. \lag\ integrated
along the boundary. For example, doing straightforward perturbation theory in
$g$ and $\bar g$, we find that the 2-point function is
$$\vev{\partial X(z) \bar \partial X(\bar w)}={S(g, \bar g)\over
(z-\bar w)^2}\ ,
$$
where the $1\to 1$ amplitude turns out to have the expansion
$S(g, \bar g)=-1+2\pi^2 g\bar g+ \ldots$. In the operator formalism,
\eqn\oto{\vev{\partial X(z) \bar \partial X(\bar w)}=
\bra{B(g, \bar g)} \partial X(z) \bar \partial X(\bar w)\ket {0}\ , }
where $\bra{B(g, \bar g)}$ is the exact state for the dynamical
boundary, which is determined implicitly by Eqn.~\partclosed.
Let us note that the only contribution to Eqn.~\oto\ comes from
the $(1, 0)$ module. The coefficient of this module in
the expansion of $\bra{B(g, \bar g)}$ can be read off
from Eqn.~\partclosed with the result
$S(g, \bar g)=1-2[\cos 2\Delta(g, \bar g)]_{g\bar g}$,
The subscript means that we are to keep only the powers of $g\bar g$ in the
perturbative expnsion of the right-hand side.

The most remarkable feature of our reflection $S$-matrix is that its $n\to m$
connected pieces, while non-trivial, are entirely determined by the $1\to 1$
amplitude. To show how this works, let us consider the $2\to 2$ amplitude,
$$\eqalign{&\vev{\partial X(z)\partial X(u)\bar \partial X(\bar w)
\bar \partial X(\bar v)}=G(z, u, \bar w, \bar v)=\cr &
{1\over (z-u)^2 (\bar w-\bar v)^2}
+{S^2\over (z-\bar v)^2 (u-\bar w)^2}+
{S^2\over (z-\bar w)^2 (u-\bar v)^2}
+F(z, u, \bar w, \bar v)\ , }
 $$
where $F$ is the connected part. The existence of a null state
among the descendants of $\partial X$ at level 3 gives rise to
a 3-rd order differential equation \BPZ,
\eqn\diffeq{\left [{\partial^3\over \partial z^3}-
4{\partial\over \partial z}\sum_{i=1}^3
{\partial\over \partial w_i}{1\over z-w_i}-
6\sum_{i=1}^3{\partial\over \partial w_i}{1\over (z-w_i)^2}\right ]
G(z, w_1, w_2, w_3)=0\ .}
This equation determines the connected part in terms of the disconnected
parts, and we find
$$F(z, u, \bar w, \bar v)={2(1-S^2)\over (z-\bar v)
(z-\bar w) (u-\bar v) (u-\bar w)}
$$
The Fourier transform of the Minkowskian continuation of this is
\eqn\Ft{\tilde F(E_1, E_2; E_3, E_4)=2(1-S^2)\delta(E_1+E_2-E_3-E_4)
(E_1+E_2-|E_1-E_3|-|E_2-E_3|)\ ,}
where $E_i>0$. Curiously, this formula bears a strong resemblance to
the $2\to 2$ amplitude found in the $c=1$ matrix model \smat.

It appears that recursive application of the differential
equations to higher-point functions determines them entirely.
We find that the connected part of any $1\to n$ amplitude vanishes,
while for the $2\to 2n$ amplitudes
$$\vev{\partial X(z_1) \partial X(z_2) \bar \partial X(\bar w_1)
\ldots \bar \partial X(\bar w_{2n})}_{\rm conn}=
-(-2)^n (1-S^2){(z_2-z_1)^{2n-2}\over \prod_{i=1}^{2n} (z_1-\bar w_i)
(z_2-\bar w_i)}\ .
$$
In Fourier space, these become piecewise linear functions of the
energies, similar to Eqn.~\Ft. This should be compared with some results
in dissipative quantum mechanics \freed.

\newsec{Conclusions}

In this Letter we have presented a solution of the $c=1$
conformal field theory with a boundary sine-Gordon interaction.
Conformal invariance imposes tight constraints on the
partition function when viewed both from the open string and
from the closed string point of view. Examination of the
\csc\ allows us to deduce the exact boundary state in terms of
a universal function of the complex potential strength.
Remarkably, the boundary state is built out of the Virasoro modules
of the well-known discrete states of the $c=1$ CFT.
{}From this information we determine a new non-trivial $S$-matrix for
the scattering of the massless scalar quanta from the dynamical
boundary. The operator $\partial X$ has a null state at level 3,
and this gives rise to BPZ differential equations which appear
to fix all of the correlations functions recursively.
Every correlator is a simple rational function of coordinate differences,
which is certainly extraordinarily simple behavior for a nontrivial
field theory. In future work we plan to expand on our results, and to discuss
their applications to specific physical systems.

\vskip .3in
\centerline{\bf Acknowledgements}
\vskip .3in

We thank J. Maldacena and A. Yegulalp for useful discussions.
The work of C.G.C. was supported in part by DOE grant DE-FG02-90ER40542
and by the Monell Foundation.
The work of I. R. K. was supported in part by DOE grant DE-AC02-84-1553,
NSF Presidential Young Investigator Grant No. PHY-9157482,
James S. McDonnell Foundation Grant No. 91-48 and the A. P. Sloan Foundation.

\listrefs
\bye